# Memristive Transfer Matrices

Blaise Mouttet

George Mason University, bmouttet@gmu.edu

## ABSTRACT

An electrical analysis is performed for a memristor crossbar array integrated with operational amplifiers including the effects of parasitic or contact resistances. It is shown that the memristor crossbar array can act as a transfer matrix for a multiple input-multiple output signal processing system. Special cases of the transfer matrix are described related to reconfigurable analog filters, waveform generators, analog computing, and pattern comparison.

## 1  INTRODUCTION

The memristor (i.e. memory resistor) is a fundamental new type of circuit element offering dynamic capabilities beyond the static characteristics of ordinary resistors. Reference to the theoretical and practical development of memristive materials is given by [1-5]. It is notable that, while the memristance effect of $TiO_2$ has attracted much recent attention based on the publications from HPLabs, original research on the memory resistance effects of $TiO_2$ was performed in the 1960's [6]. In addition, numerous other materials have long been known to exhibit memory resistance effects [7], although the connection to the later developed memristor theory was not realized until recently. It is also worth noting that a similarly named "memistor" was developed by Bernard Widrow and Ted Hoff in the 1960's as a component of one of the first implementations of a neural network named ADALINE [8]. However, the Widrow-Hoff memistor was a 3-terminal device as opposed to the 2-terminal memristor and was implemented using an electroplating cell rather than solid state circuitry compatible with integrated circuit fabrication.

The most common applications suggested for memristors have so far been in the areas of non-volatile memory and neuromorphics. However, the market of non-volatile memory is well-developed with existing solutions that continue scaling despite the continual prediction of the end to Moore's Law. In addition, many other more fully developed technologies such as phase change memory, FeRAM, and MRAM are also competing for non-volatile memory market share which will make it very difficult for a memristor solution to compete.

Memristive neuromorphics offers more promise as an emerging technology but, lacking an association with any existing market need, it has an unclear present value. More conventional analog electronics applications may offer an easier entry point for memristors into commercial products. There has been some analysis given toward general analog applications of individual memristors [9,10] as well as discussion of analog applications of memristor arrays arranged in a crossbar configuration [11,12]. However, there has been little analysis of memristor crossbar arrays integrated with analog electronics. Section 2 describes a design approach to analog circuitry based on memristor arrays in a crossbar configuration integrated with operational amplifiers. Section 3 includes an electrical analysis of the memristive transfer matrix. Section 4 describes special cases of the memristor transfer matrix for specific applications.

## 2  MEMRISTOR TRANSFER MATRIX

In modern control systems, transfer matrices are used to define a functional relationship between a set of input signals $u_i(t)$ ($0 \leq i \leq n$) and a set of output signals $y_j(t)$ ($0 \leq j \leq m$). For linear systems (or linear approximations of non-linear systems) this relationship may be written in the Laplace domain in terms of a matrix transformation:

$$Y_j(s) = \sum_i T_{ij}(s) U_i(s) \qquad (1)$$

where $T_{ij}(s)$ represents transfer matrix elements. For systems having a memory dependence the elements $T_{ij}$ need to reflect this memory dependence. When building circuits using circuit elements such as resistors, capacitors, and amplifiers such memory effects are not readily achievable. Software controlled processors are more conventionally used to implement such a transfer matrix and can provide both adaptability and memory to the elements $T_{ij}$. However, for large matrices high operational speeds are required to simultaneously update the matrix elements and calculate new output signals resulting in a limit to real time response rates. This is in sharp contrast with the observed rate of firing of neurons in the human brain (<100Hz) which perform pattern recognition and real-time response in a fashion far superior to any known computer. The implementation of (1) using memristive junctions could provide memory to the elements $T_{ij}$ thus integrating the adaptability of software solutions with the real time response of hardware.

For the purposes of this discussion a memristive junction is used to refer to a voltage-controlled memristive system junction defined by a generalized version of Ohm's law relating a voltage signal v(t) and a current signal i(t) such that:

$$i(t) = v(t)/R(w(t)) \qquad (2)$$

where $R(w(t))$ is a memristance function dependent upon a state parameter $w(t)$ defined in accordance with a differential equation having the form:

$$\frac{dw(t)}{dt} = f(w(t), v(t)) \qquad (3)$$

In terms of thin film materials the function $f(w(t),v(t))$ may either represent the voltage dependence of ionic or oxygen vacancy drift or the growth rate of metallic filaments in a thin film junction. Such functional dependence has been shown to obey an exponential dependence by [4] and, while high electric fields produce fast switching, small signals applied for short durations produce negligible alteration the state parameter (i.e. dw/dt=0). Thus for small signals the memristive junction may be approximated as a linear resistor having a value dependent on the duration and amplitude of a previously applied large voltage signal.

Fig.1 illustrates a memory cell designed for applying both programming voltages and small signal voltages to a memristive junction. A memristive junction is indicated by an impedance value $Z_{ij}(w)$ accounting for capacitance as well as the programmable resistance of the junction. As suggested by [10] a p-type MOSFET is included to controllably decrease the resistance of the memristive junction while an n-type MOSFET is included to controllably increase the resistance. When both transistors are in the OFF state a small signal current may be applied via diode D. The diode is included in the memory cell to prevent sneak current paths when incorporated into a crossbar architecture.

Fig. 2 illustrates a 2x2 signal processing matrix formed using the memristive memory cells arranged in a crossbar configuration and including operational amplifiers connected to the outputs. Parasitic impedances at the input to the crossbar are denoted by Zp,in(1) and Zp,in(2) and parasitic impedances at the output of the crossbar are denoted by Zp,out(1) and Zp,out(2). The effects of such parasitics may be more substantial at the nanoscale in which defects and non-negligible contact resistances become significant. Row selection transistors are provided to control the connection of memristive junctions to ground during the programming of the memristive state. The operational amplifiers include feedback impedances indicated by Zf,out(1) and Zf,out(2).

## 3 ELECTRICAL ANALYSIS

Fig. 3 illustrates the electrical diagram of Fig. 2 in which the operational amplifiers are represented in terms of their input impedances $Z_{op,in}$, output impedances $Z_{op,out}$, and voltage gains $A_v$. An approximate linear system analysis may be performed assuming that the DC level of the input voltages $V_{in}$ are greater than the diode thresholds $V_{th}$ and that $V_{in}-V_{th}$ has a sufficiently small magnitude and is applied for a sufficiently short time so as to avoid memristance drift. In this case the superposition theorem may be applied and the total output signal voltages may be calculated as the sum of the partial output voltages produced from the individual input voltage signals with the other input signal grounded. The relationship between column wire voltage $V_c$ and input voltage $V_{in1}$ may be calculated based on Kirchhoff's current law applied to the column wire (assuming zero current from the reverse biased diodes $D_{21}$ and $D_{22}$) producing:

$$\frac{V_{in1} - V_c}{Z_{p,in1}} = \frac{V_c - V_{th11}}{Z_{11}(w) + Z_{p,out1} + Z_{op,in1}||(Z_{F1} + Z_{op,out1})} + \frac{V_c - V_{th12}}{Z_{12}(w) + Z_{p,out2} + Z_{op,in2}||(Z_{F2} + Z_{op,out2})} \qquad (4)$$

from which it is straightforward to calculate $V_c$ in terms of $V_{in1}$.

For the more general case of a crossbar including N columns and M rows the relationship between the input voltages and the column wire voltages may be expressed as:

$$V_{in,i} - V_{c,i} = \sum_j \alpha_{ij}(s)(V_{c,i} - V_{th,ij}) \qquad (5)$$

where $\alpha_{ij}$ is given by:

$$\alpha_{ij} = \frac{Z_{p,ini}}{Z_{ij}(w) + Z_{p,outj} + Z_{op,inj}||(Z_{Fj} + Z_{op,outj})} \qquad (6)$$

and i and j are indices ranging from $1 \leq i \leq N$ and $1 \leq j \leq M$.

Back to the 2x2 case of Fig. 3 an application of Kirchhoff's voltage law from $V_{c1}$ on row 1 produces two coupled equations:

$$V_{c1} - V_{th11} - i_1(Z_{11}(w) + Z_{p,out1}) - (i_1 - i_2)Z_{op,in1} = 0 \qquad (7)$$

$$(i_1 - i_2)Z_{op,in1} - i_2(Z_{f1} + Z_{op,out1}) - A_v(vp - vn) = 0 \qquad (8)$$

where $i_1$ is the current flowing through the memristive junction, $i_2$ is the current flowing through the feedback impedance of the operational amplifier connected to row 1, and vp-vn is the differential voltage between the non-inverting and inverting input terminals of the operational amplifier which is amplified by $A_v$ at the output. The differential voltage may also be expressed in terms of the current as:

$$vp - vn = (i_1 - i_2)Z_{op,in1}. \quad (9)$$

and the output voltage Vout1 may be calculated based on the amplified differential voltage.

The result of this analysis produces an amplified differential voltage equal to:

$$V_{out} = \frac{Z_{f,1} + Z_{op,out1}}{\frac{1-A_v}{A_v} + \frac{Z_{f1} + Z_{op,out1}}{A_v Z_{op,in1}}} \times \quad (10)$$

$$\frac{V_{c,1} - V_{th11}}{Z_{11}(w) + Z_{p,out1} + Z_{op,in1} - \frac{Z_{op,in1}^2(1-A_v)}{Z_{op,in1}(1-A_v) + Z_{f1} + Z_{op,out1}}}$$

For an ideal operational amplifier in which Zop,in1=∞, $A_v$ =∞, and Zop,out1=0 this simplifies to:

$$V_{out1} = -\frac{Z_{f,1}}{Z_{11}(w) + Z_{p,out1}}(V_{c,1} - V_{th,11}). \quad (11)$$

Applying the superposition theorem for a MxN crossbar (11) may be generalized to find the total output voltage signals based on all of the input signals applied simultaneously:

$$V_{out,j} = -\sum_i \frac{Z_{f,j}}{Z_{ij}(w) + Z_{p,outj}}(V_{c,i} - V_{th,ij}). \quad (12)$$

where Voutj is the operational amplifier output from the jth row. In cases where the coefficients $\alpha_{ij}$ from (6) are negligible (which may be designed by increasing the sum of the crossbar output impedances and op-amp feedback impedances relative to the crossbar input impedances) the column voltages reduce to the input voltages and (12) is representative of a signal transfer matrix as in (1) where the memristive resistance states Zij(w) effectively determine the magnitude of the elements of the transfer matrix.

## 4 APPLICATIONS

Several possible applications of memristive transfer matrices were discussed at the 1st Memristor and Memristive Systems Symposium [13]. Depending on the resistive and/or capacitive values of the fixed impedances different functionalities can be achieved as explained below.

### 4.1 Programmable Analog Filters

In the case where the impedances are chosen such that $\alpha_{ij}$ are negligibly small (e.g. |Zp,ini|<<|Zp,outj+Z,Fj|), the output impedance of the crossbar are purely resistive (Zp,outj=Rp,outj), and the feedback impedances $Z_{F,j}$ of the operational amplifiers are implemented by resistors $R_{F,j}$ in parallel with capacitors $C_{F,j}$ Eq.(12) expressed in the Laplace domain reduces to:

$$V_{out,j}(s) = -\sum_i \frac{R_{F,j}}{(1 + sR_{F,j}C_{F,j})(Z_{ij}(w) + R_{p,outj})}V_{in,i}(s) \quad (13)$$

where Vin,i(s) represents the LaPlace Transform of the small signal inputs and Vout,j(s) represents the Laplace Transform of the small signal outputs. The transfer matrix elements of (13) are representative of a low pass filter having a tunable gain dependent upon the memristive states Zij(w).

In a similar case to that above but in which the output impedance of the crossbar are implemented by capacitors (Zp,outj=Cp,outj) and the feedback impedances $Z_{f,j}$ of the operational amplifiers are implemented by resistors $R_{F,j}$ Eq.(12) may be expressed in the Laplace domain as:

$$V_{out,j}(s) = -\sum_i \frac{sR_{F,j}C_{p,outj}}{sZ_{ij}(w)C_{p,outj} + 1}V_{in,i}(s) \quad (14)$$

The transfer matrix elements of (14) are representative of a high pass filter having a tunable half-power frequency dependent upon the memristive states Zij(w). By cascading multiple memristive transfer matrices which perform the operations of (13) and (14) bandpass filters may be constructed in which both amplitude and frequency tuning can be performed.

### 4.2 Waveform Generators

If the signal voltage inputs to the columns wires of the memristive transfer matrix include a relative delay period T between adjacent rows and resistors (Rf,j, Rp,outj) are used for the feedback impedance and crossbar output impedance (12) reduces to:

$$v_{out,j}(t) = -\sum_i \frac{R_{f,j}}{Z_{ij}(w) + R_{p,outj}}v_{in}(t - iT). \quad (15)$$

For this implementation the memristive states can be used to determine modified pulse trains based on the basic pulse train in which the amplitude of pulses are modified based on the magnitude of Zij(w). In a similar case in which the input signals are harmonics of a sine wave having an angular frequency ω (12) reduces to:

$$v_{out,j}(t) = -\sum_i \frac{R_{f,j}}{Z_{ij}(w) + R_{p,outj}} \sin(i\omega t). \quad (16)$$

In this implementation tuning of the memristance states may be used in the generation of approximate Fourier series representations of periodic signals.

### 4.3 Analog Arithmetic

For the purposes of analog arithmetic the high resistance state of a memristive junction may represent a logic 0 in which very little current is passed while the low resistance state of the memristance junction may be represent a logic 1. In this case each column of the crossbar may be interpreted as storing a binary string representative of a numerical value and selection of the input voltages to the columns may be used to control a summation operation. In order to separate the crossbar rows from a least significant value to a most significant value precision resistors at the outputs of the crossbar may be set to multiples of 2 times the feedback resistance (Rp,outj = $2^j$Rf,j). In this case (12) reduces to:

$$v_{out,j}(t) = -\sum_i \frac{1}{Z_{ij}(w)/R_{f,j} + 2^j} v_{in,i}(t). \quad (16)$$

This implementation produces analog output signals having a magnitude in proportion to the summation of the memristive states of the crossbar rows. Further discussion of this approach to analog arithmetic is found in [14].

### 4.4 Pattern Comparison

In terms of binary logic the comparison between two individual bits (A,B) is performed using the XNOR function:

$$X = AB + \overline{A}\overline{B} \quad (17)$$

where X = 1 when A and B are identical and X = 0 when A and B are different. A more generalized comparison function may be defined between M strings of binary data $A_{ij}$ wherein each string has a length N and a string of binary data $B_i$ having a length N where 1< i <N and 1<j<M:

$$X_j = \sum_i (A_{ij}B_i + \overline{A}_{ij}\overline{B}_i). \quad (18)$$

In this case $X_j$ is an integer ranging from 0 to N depending upon the number of bits in $A_i$ and $B_i$ which are identical.

The function of (18) may be implemented using two memristive transfer matrices each performing the function (12) but with complementary resistance states in which bit patterns $A_{ij}$ and its complement (not $A_{ij}$) are implemented by the memristive junctions set to the maximum resistance $R_{HIGH}$ (logic 0) or minimum resistance $R_{LOW}$ (logic 1) and $B_i$ are implemented by the binary voltage states of the input signals and the equivalent inverted pattern (not $B_i$). Further discussion of this implementation is found in [15].

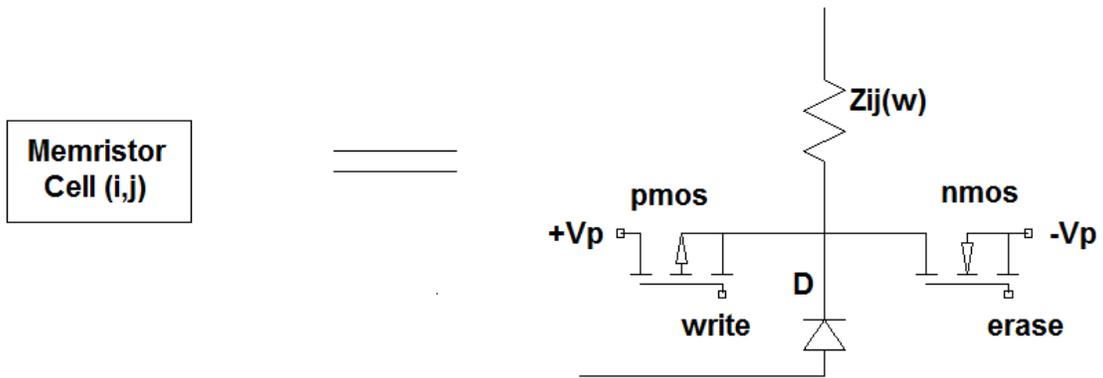

Fig.1. Memristor memory cell.

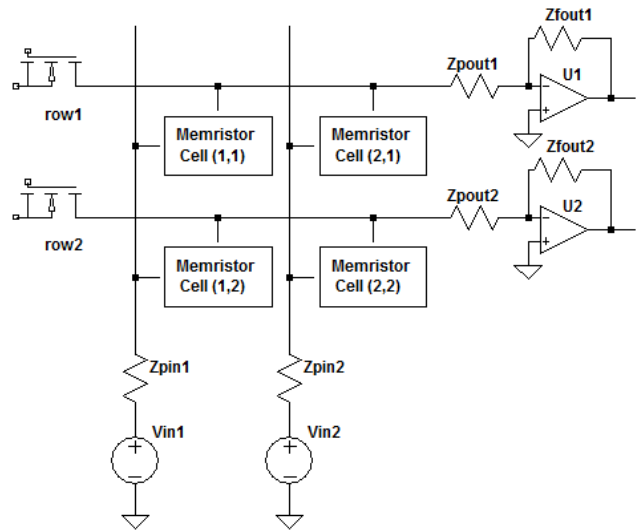

Fig.2. 2x2 Memristor signal transfer matrix.

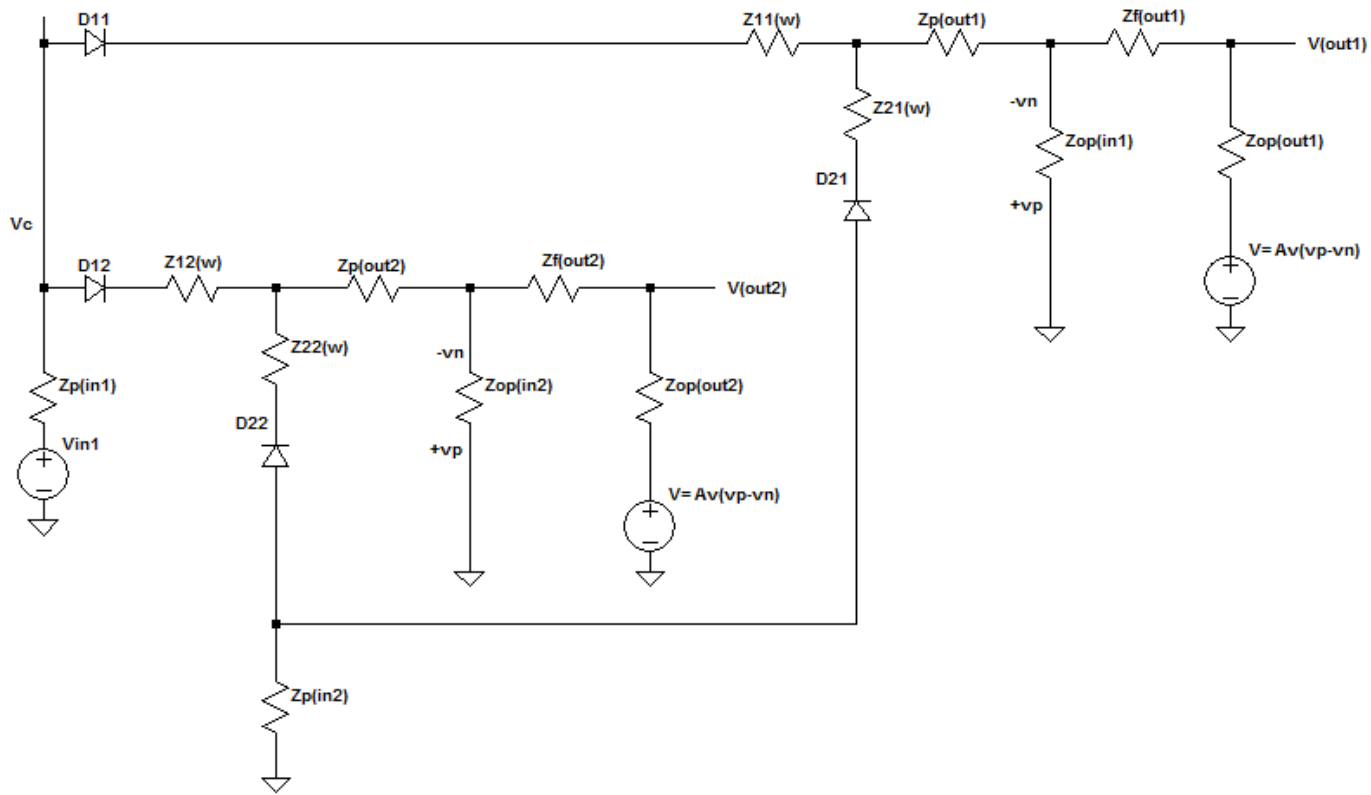

Fig.3. Electrical diagram of 2x2 Memristor signal transfer matrix with Vin2=0.